# Demand Response by Aggregates of Domestic Water Heaters with Adaptive Model Predictive Control

Francesco Conte
Faculty of Engineering
Campus Bio-Medico University of Rome
Roma, Italy
f.conte@unicampus.it

Stefano Massucco, Federico Silvestro
DITEN
University of Genova
Genova, Italy
stefano.massucco@unige.it

Diego Cirio, Marco Rapizza
Ricerca sul Sistema Energetico
RSE S.p.A.
Milano, Italy
diego.cirio@rse-web.it

*Abstract*— This paper describes an intelligent management algorithm for an aggregate of domestic electric water heaters called to provide a demand response service. This algorithm is developed using Model Predictive Control. The model of the entire aggregate is dynamically identified using a recursive polynomial model estimation technique. This allows the control to be adaptive, *i.e.,* able to adjust its decisions to the system characteristics, which vary over time due to the daily distribution of users' hot water consumption. To answer the demand response requirements, aggregated power variations are realized by modifying the temperature set-points of the water heaters without compromising the users' comfort. The developed approach allows tracking a regulation signal and mitigating the so-called rebound, *i.e.,* the recovery of energy needed by the aggregate at the end of the service to return to the baseline thermal state. Analyses in a simulation environment allow the validation of the potentialities of the proposed method.

## I. Introduction

Integrating non-programmable Renewable Energy Sources (RESs), required by decarbonization goals, poses significant challenges for managing the electrical power system, especially those related to frequency regulation. The strategies adopted to address this issue are mainly based on flexibility: devices that in the past played a "rigid" role, such as the same RESs and loads, are now involved in providing system regulation services.

The provision of these services by loads is called Demand Response (DR). The idea is that loads modulate their power/energy consumption, responding to some command of the Transmission System Operator (TSO). Such modulation is possible only if loads are provided with some form of energy reserve to vary their power/energy consumption without compromising the end-user specifications. Loads related to temperature control, such as air cooling and heating systems, fridges, and Electric Water Heaters (EWHs), are clear examples of flexible loads since they keep a thermal energy reserve.

To realize a DR service is however complex for multiple reasons. Indeed, loads are numerous and highly distributed over the network, making them difficult to control and monitor. Moreover, they are significantly smaller than traditional generation units, meaning that many of them should be used to get a consistent regulation service. The solution to these issues is to constitute aggregates of loads. In general, an aggregate is a set of generation units or loads that interface with the grid and the service market as a unique entity, monitored and managed by an aggregator.

Following the European guidelines [1]-[2], deliberation [3] of the Italian Regulatory Authority for Energy, Networks, and Environment has led the Italian TSO to write the codes of pilot projects for the "enlargement of the dispatching resources." In these projects, aggregated entities, called Virtual Qualified Units (VQUs), can be composed of loads [4], generating units, or a mix of them [5]. The dispatching and balancing service designed in [5] asks for a variation from a baseline profile of the power generation or demand of the VQU, for a given time extension, from a minimum of 15 minutes up to 3 hours.

In [6], the capability of domestic thermostatically controlled EWHs to contribute to this service as a part of a VQU is evaluated. More specifically, by the method in [6], it is possible to establish and predict a baseline profile of the power consumption of an aggregate of EWHs and an upward and a downward margin, within which the consumption profile of the aggregate can be modulated, without compromising the users' comfort. This is obtained from a Monte Carlo analysis, which uses the thermal model of EHWs and properly elaborates the external inputs, such as ambient and cold water temperatures, and hot water demand.

The objective of the present work is to develop an algorithm able to coordinate the response of the single EWHs to realize the mentioned DR service based on the flexibility margins estimated by the approach introduced in [6].

Literature provides many solutions to control the power consumption of aggregates of EWHs or, more in general, of thermal loads. For example, in [7]-[8], aggregates of buildings' temperature control systems are managed, through a neural network and reinforcement learning methods, to respond to price signals, assuming the possibility of controlling the on/off state of the single devices. A similar approach is applied in [9] to thermostatically controlled EWHs. The power consumption profile of aggregates of domestic EWHs is also controlled in [10] and [11]. In these two papers, the control is operated both by modifying the temperature set-points and the on/off state of the single devices.

This work has been financed by the Research Fund for the Italian Electrical System in compliance with the Decree of Minister of Economic Development April 16, 2018.

Performances of the literature approaches are very promising. However, many commercial EWHs do not allow remote control of their on/off state. Moreover, many solutions assume to know the thermal state of each EWH, and accordingly, they decide their individual contribution. Instead, our objective is to propose a simple control algorithm that observes only the power consumption of the aggregate and controls only the temperature set-points of the EWHs.

The method proposed in Section II is based on a Model Predictive Control (MPC) that uses a simple dynamical model of the aggregated power consumption. This model is time-varying and identified online with a recursive polynomial model estimation technique [12]. Simulation results reported in Section IV show that the aggregate can be accurately controlled by adopting this simple method, which broadcasts a unique temperature set-point variation as the control signal to all the EWHs.

## II. Proposed Methodology

### A. Problem forumlation

We first introduce the problem hypotheses. Given a set of $N$ EWHs: i) the $i$-th EWH ($i = 1,2,...,N$) is controlled by a standard thermostatic logic with two temperature thresholds: $\theta_i^* - \Delta_i$ and $\theta_i^* + \Delta_i$, where $\theta_i^*$ is the temperature set-point and $2\Delta_i$ is the thermostat dead-band extension; ii) the available remote control signal is the variation of the temperature set-points $\Delta\theta_i, i = 1,2,...,N$; iii) the total active power consumed by the aggregate $P_a$ is measured; iv) measurements and control signals are sampled with a granularity of 1 minute.

The control objective is to follow a reference signal $P_a^*$, which is (according to the rules designed in [5]):
  a) equal to the baseline $P_a^b$, when the DR service is not active;
  b) equal to $P_a^b + \Delta P_a^*$, when the DR service is active; $\Delta P_a^*$ is a desired variation, required by the TSO, to be kept for a time interval $\Delta t$ and starting from time $\tau$.

The transition from $P_a^* = P_a^b$ to $P_a^* = P_a^b + \Delta P_a^*$ and viceversa is realized with a linear ramp lasting 5 minutes.

### B. Control solution

The control architecture is depicted in Fig. 1. The active power consumed by the aggregate $P_a$ is represented by the following linear discrete-time model:

$$P_a(k+1) = a(k)P_a(k) + b(k)\Delta\theta(k) + w(k) \quad (1)$$

where: $\Delta\theta(k)$ is the unique the set-point variation, sent to all the EWHs within the aggregate; $a(k)$, $b(k)$ and $w(k)$ are the time-varying coefficients of the model. More specifically: $a(k)$ represents the time correlation between the consumed power at time step $k$ and the one at the next time step $k+1$; $b(k)$ models the effect of the set-point variation $\Delta\theta(k)$ on the aggregate power consumption; $w(k)$ is a disturbance term, which models all the externalities, *i.e.*, the ambient temperature and the water use. These three parameters are dynamically identified by the *Model Identifier*, which only uses the measurements of the aggregated power $P_a$ and of the set-point variation $\Delta\theta$.

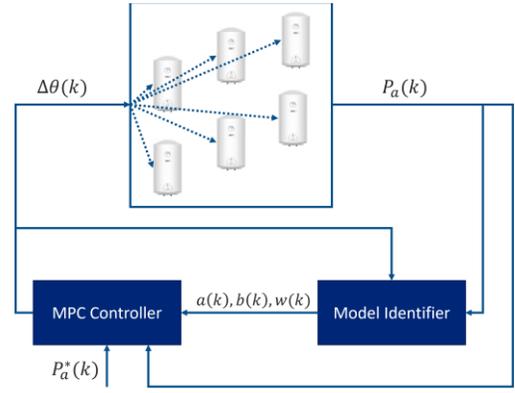

Fig. 1. Control architecture.

Given the simplified aggregate model (1), the objective is to define the control signal $\Delta\theta(k)$ based on an optimization policy by which the distance $|P_a(k) - P_a^*(k)|$ is minimized. This task is carried out by the *MPC Controller*.

In the following, we detail the algorithms implemented on the two subsystems, *Model Identifier* and *MPC Controller*.

#### 1) Model Identifier

The model parameters are identified by a method of recursive estimation of polynomial dynamical models, a class into which model (1) falls. The adopted method uses an AutoRegressive eXogenous (ARX) model structure and a Kalman filter with infinite memory as a recursive estimator. The details of this technique can be found in [12].

#### 2) MPC Controller

The adopted control method is MPC. This technique consists of computing, at any time step $k$, the solution of a constrained optimal control problem, defined over a time interval from the current time $k$ to the so-called time horizon $k + T$. The optimal control problem provides a prediction of the time evolution of the controlled system by including the system dynamical model in the constraints set. Operation and technical constraints, such as state and control variable limits, can also be added to the problem. The solution of the optimization problem returns an optimal control trajectory $\{u^*(\tau)\}_{\tau=k}^{k+T-1}$. The first element of this trajectory $u^*(k)$ is applied at time $k$. Then, at the subsequent time step $k + 1$, the procedure is repeated based on an updated version of the optimization problem. This strategy is known as the *receding horizon principle*, which allows MPC to be a *closed-loop* control method conferring it excellent robustness against model and prediction errors.

The receding horizon principle also allows MPC to be easily transformed into an *adaptive* control method. Indeed, if the controlled system is stationary (with constant parameters), the optimization problem update, occurring at any time step $k$, only consists of the update of the current system state, based on new measurements. If the system is time-varying (with time-varying parameters), the optimization problem update can easily include the updated values of the model parameters.

The specific optimal control problem formulated for the control of the EWHs aggregate is the following:

$$\{\Delta\theta^*(j)\}_{j=k}^{k+T-1} = \arg\min_{\{\Delta\theta(j)\}} J \quad (2)$$

$$J = \sum_{j=k}^{k+T-1}\left[w_P\big(P_a(j)-P_a^*(j)\big)^2 + w_\theta\big(\Delta\theta(j)\big)^2 \right.$$
$$\left. + w_{d\theta}\big(\Delta\theta(j)-\Delta\theta(j-1)\big)^2\right] \quad (3)$$
$$+ w_P\big(P_a(T)-P_a^*(T)\big)^2$$

such that, for all $j = k, k+1, \dots, k+T-1$,

$$P_a(j+1) = a(k)P_a(j) + b(k)\Delta\theta(j) + w(k) \quad (4)$$

$$\Delta\theta^{min} \leq \Delta\theta(j) \leq \Delta\theta^{max} \quad (5)$$

and, for all $j = k+L, k+L+1, \dots, k+T-1$,

$$\Delta\theta(j) = \Delta\theta(k+L-1) \quad (6)$$

where: $w_P$, $w_\theta$ and $w_{d\theta}$ are the optimization weights, associated with the tracking of the reference signal, to the magnitude of the control signal $\Delta\theta$, and to the variation of the control signal, respectively; $\Delta\theta^{min}$ and $\Delta\theta^{max}$ are the limits of the control signal, i.e., the maximum and minimum allowed variations of the temperature set-point; $T$ is the prediction horizon; $L \leq T$ is the control horizon: the control trajectory is indeed decided up to the time $k+L-1$ and supposed to be constantly kept equal to the last decided value for time $k+L-1$, as imposed by constraint (6).

Constraint (4) is the dynamical model of the aggregate: thanks to it, the controller will predict the time evolution of the aggregated power $P_a$ along the prediction time interval and decide the optimal control trajectory $\{\Delta\theta^*(j)\}_{j=k}^{k+T-1}$. Notice that model in (4) uses parameters $a(k)$, $b(k)$ and $w(k)$ updated at time $k$ and assumed to be constant along the prediction interval. As explained before, the dynamical update of these parameters makes the controller *adaptive*. In the case of EWHs, this is crucial since the variation of the ambient temperature and, mainly, of the water use significantly modify the effect of the temperature set-point variation on the aggregate power consumption during the day. $P_a(k)$ and $\Delta\theta(k-1)$ are known terms of the optimization problem since by hypothesis iii) (given in Section II.A), at time step $k$, the measurement of $P_a$ is available, whereas $\Delta\theta(k-1)$ is the last set-point variation computed by the same controller.

Weights $w_P$, $w_\theta$ and $w_{d\theta}$ are control parameters. Their values define the priority among the accuracy in tracking the reference signal $P_a^*$, the limitation of the temperature set-point variation $\Delta\theta$, and its time variation. Therefore, if $w_P \gg w_\theta, w_{d\theta}$, the controller will prioritize the tracking of the reference signal, accepting that $\Delta\theta$ and its time variation are arbitrarily large, within limits established by constraint (5). Whereas, if the weights values are comparable to each others, the controller will find a solution that reduces as much as possible both the tracking error and the magnitude and the variation of the control signal.

Therefore, two control modes (CMs) are defined. In CM 1, optimization weights have comparable values. This CM is used when the DR service is off, to follow the baseline. Indeed, during this phase, users' choices should be preserved; thus, their personal temperature set-points are varied as less as possible. This can be easily achieved since the baseline profile has been computed based on the typical set-point choices of users. In CM2, $w_P \gg w_\theta, w_{d\theta}$. This CM is used when the DR service is active and, thus, to track the reference signal is a priority. In this phase, the variation of the users set-points is more acceptable by virtue of the request to deliver the DR service. The transition between the two CMs is realized by a linear ramp variation of the optimization weights within a time interval of 5 minutes.

At the end of the service delivery time interval $\Delta t$, two control options can be applied: 1) *Immediate switching to CM 1*. In this way, the controller comes back to penalize the magnitude and the set-point variation similar to the inaccuracy in tracking the aggregated power reference signal, which is back to being the baseline. This will cause an immediate energy recovery, called *rebound*, i.e., a variation of the power aggregated power profile with a sign opposite to the one realized to deliver the DR service. 2) *Rebound mitigation*. Instead of immediately switch from CM 2 to CM 1, CM 2 is kept for a time interval equal to $\Delta t$, and the switch to CM 1 is then carried out with a linear ramp transition during 1 hour. Fig. 2 graphically reports the two CMs scheduling options, with or without the rebound mitigation.

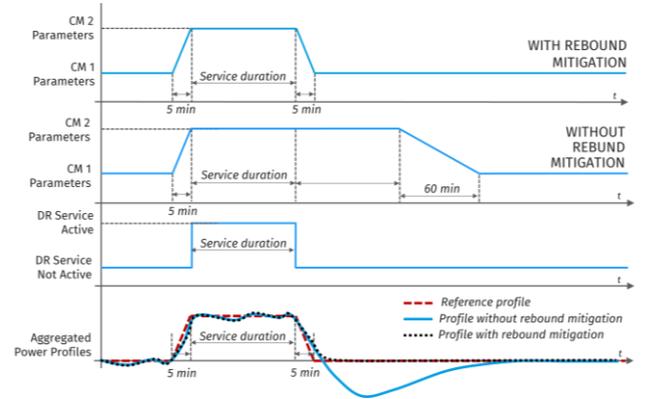

Fig. 2. Time scheduling of the CMs.

## III. SIMULATIONS AND RESULTS

The adaptive MPC controller has been implemented on the Matlab/Simulink platform. The adopted control parameters are reported in Table I. Two aggregates of EWHs are considered. They are both composed of 600 DHWs belonging to three classes of commercial models, whose parameters are reported in Table II. *Aggregate 1* is only made by the 100-liters EHWs, and the total nominal power is $P_a^{nom} = 900$ kW; *Aggregate 2* is made by the three classes of EWHs partitioned with the percentages reported in Table II and the total nominal power is $P_a^{nom} = 752$ kW. Both the aggregates are supposed to be localized in the Italian climate zone D (see [6] for details). Their power consumption is simulated by taking into account the ambient temperature, the cold water temperature, and the hot water use, as detailed in [6], during a typical day in May. It is worth mentioning here that simulations have been carried out assuming that the aggregates are localized in climate zones different from D and in months different from May, with performances comparable to the ones shown in the following.

TABLE I. ADAPTIVE MPC CONTROLLER PARAMETERS

| Description | Symbol | Value* | Unit |
|---|---|---|---|
| Reference signal tracking weight CM 1 | $w_P$ | $50/P_a^{nom}$ | kW$^{-1}$ |
| Reference signal tracking weight CM 2 | $w_P$ | $5/P_a^{nom}$ | kW$^{-1}$ |
| Temperature set-point variation weight CM 1 | $w_\theta$ | 0.5 | °C$^{-1}$ |
| Temperature set-point variation weight CM 2 | $w_\theta$ | 0.01 | °C$^{-1}$ |
| Control signal variation weight CM 1 | $w_{d\theta}$ | 1 | °C$^{-1}$ |
| Control signal variation weight CM 2 | $w_{d\theta}$ | 0.001 | °C$^{-1}$ |
| Prediction horizon | $T$ | 30 | min |
| Control horizon | $L$ | 5 | min |

*$P_a^{nom}$ is the nominal power of the controlled DHW aggregate.

TABLE II. REFERENCE MODELS OF DOMESTIC EWHs IN ITALY

| Model | ARISTON PRO ECO R 50 V/3 | ARISTON PRO ECO R 80 V/3 | ARISTON PRO ECO R 100 V/3 |
|---|---|---|---|
| Capacity ($V$) | 50 l | 80 l | 100 l |
| Nominal Power ($P^n$) | 1.2 kW | 1.2 kW | 1.5 kW |
| Maximal temperature ($T^{max}$) | 75 °C | 75 °C | 75 °C |
| Thermal dispersion at 65°C | 0.99 kWh/d | 1.35 kWh/d | 1.56 kWh/d |
| Thermostat dead-band (2Δ) | 5 °C | 5 °C | 5° C |
| Diffusion rate [6] | 22 % | 60 % | 18 % |

TABLE III. SIMULATION SCENARIOS

| Name | Aggregate | Rebound mitigation | $\Delta P_a^*$ | DR Time slots |
|---|---|---|---|---|
| Simulation 1 | 1 | YES | 100% · $\Delta P_a^-$ | 2-4 p.m. |
| Simulation 2 | 1 | NO | 100% · $\Delta P_a^-$ | 2-4 p.m. |
| Simulation 3 | 1 | YES | 100% · $\Delta P_a^-$ | 10-12 a.m. 2-4 p.m. 8-10 p.m. |
| Simulation 4 | 2 | YES | 100% · $\Delta P_a^*$ | 3-6 p.m. |

TABLE IV. NUMERICAL RESULTS

| Simulation | MAPE [%] | APE$_{max}$ [%] | $f_{5\%}$ [%] |
|---|---|---|---|
| 1 | 0.43 | 4.85 | 0 |
| 2 | 0.81 | 8.50 | 1.77 |
| 3 | 0.54 | 5.12 | 0.57 |
| 4 | 0.40 | 3.76 | 0 |

The reference signal $P_a^*$ is computed as described in Section II.A. The required variation $\Delta P_a^*$ is expressed as a percentage of the flexibility margins $\Delta P_a^+$ and $\Delta P_a^-$. According to [6] and to the TSO rules [5], these quantities are defined as the maximum positive and negative power consumption variations achievable starting at time $\tau$ for the time interval $\Delta t$. They are estimated using the method in [6], which forecasts the baseline profile and two further profiles, one obtained with the maximum temperature set-point (upward profile) and one obtained with the user-acceptable minimum temperature set-points (downward profile). Flexibilities $\Delta P_a^+(\tau, \Delta t)$ and $\Delta P_a^-(\tau, \Delta t)$ are then defined as the minimum distance form the baseline of the upward and downward profiles within the time interval starting at time $\tau$ and lasting $\Delta t$, respectively.

We present the results of four simulations, whose scenarios are summarized in Table III. For each simulation, the Absolute Percentage Error (APE) in tracking the reference signal $P_a^*$ is computed as

$$APE(k) = \frac{|P_a(k) - P_a^*(k)|}{P_a^*(k)} \cdot 100\% \qquad (7)$$

and Table IV reports: its mean value, known as Mean Absolute Percentage Error (MAPE), its maximum value APE$_{max}$, and the percentage of times it exceeds 5%, denoted as $f_{5\%}$.

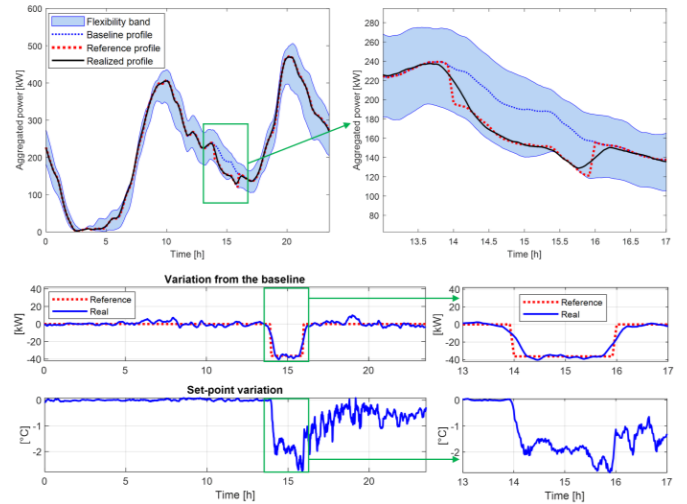

Fig. 3. Results of Simulation 1: power profile (top); variation from the baseline profile (middle); set-point variation (bottom).

Fig. 3 depicts the results of Simulation 1. The top plot shows the aggregated power consumption profile (solid black line), together with the reference signal $P_a^*$ (red dashed line), the baseline profile (blue dotted line), and the upward and downward profiles computed as detailed in [6]. We can observe that the real profile follows with good accuracy the baseline when the DR service is not active and then deviates from it between 2 and 4 p.m., chasing the reference signal. As shown in the zoom, this reference signal replicates the baseline shifted down by an amount equal to the minimum value of the flexibility margin available within the DR service time interval. The middle figure shows the variation of the realized profile from the baseline, compared with the requested variation. Thanks to this plot, we can appreciate the accuracy obtained in following the reference signal. We also note that at the beginning and the end of the DR service, the aggregate realizes transient activation and deactivation ramps. The duration of this transient is of about 15 minutes. Referring to Table IV, we note that the MAPE in Simulation 1 is 0.43 %, confirming the good accuracy in following the reference signal. The maximum APE is 4.85 %, reached during the transient, whereas the limit of 5 % is never violated. It is worth remarking that, except for transients, the APE is always lower than 2 %.

The bottom plot in Fig. 3 reports the control signal, *i.e.*, the temperature set-point variation. It is close to zero when the DR service is not active and reaches about -2.8 °C when the service is active. Notice first that this variation is sufficiently limited to guarantee the users thermal comfort. Secondly, we observe that the set-point variation is not immediately reduced at the end of the DR service delivery. This happens because in Simulation 1 we are adopting the rebound mitigation option. It is indeed important to remark that, observing the top and middle figures, there is no rebound at the end of the DR service.

Fig. 4 shows the results of Simulation 2, which is equal to Simulation 1 but without adopting the rebound mitigation option. Both in the top and in the middle figures, it is possible to observe that, at the end of the DR service, the aggregated power profile realizes a variation from the baseline with a sign opposite to the one realized to deliver the service.

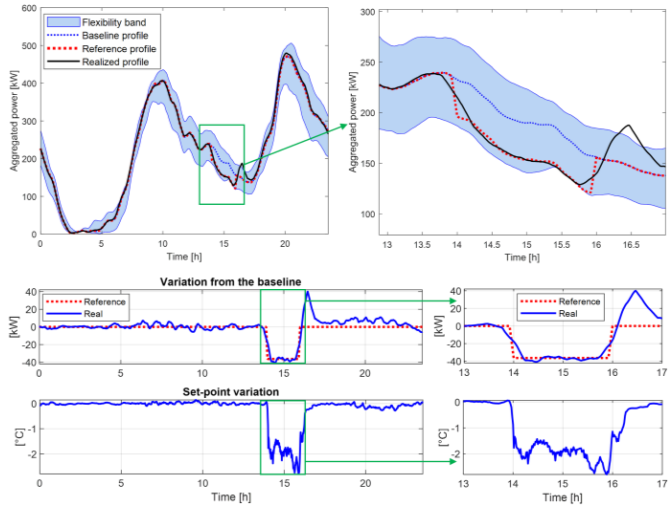

Fig. 4. Results of Simulation 2: power profile (top); variation from the baseline profile (middle); set-point variation (bottom).

This deviation is the rebound, which has a magnitude comparable to the DR one and a duration of about 1 hour. This happens because, without applying the rebound mitigation, the temperature set-point variation, depicted in the bottom plot of Fig. 4, comes back close to zero immediately after the end of the DR service. These results, compared with the ones obtained in Simulation 1 (Fig. 3), prove the effectiveness of the rebound mitigation method. Numerical indices in Table IV confirm such a conclusion. Indeed, in Simulation 2, both MAPE and maximum APE are roughly doubled with respect to Simulation 1; moreover, the 5 % limit is violated within 1.77 % of the measurement samples.

Fig. 5 and Fig. 6 show the power profiles obtained in Simulation 3 and 4, respectively. Simulation 3 aims at testing the delivery of multiple DR services on the same day with the same aggregate of Simulation 1 (Aggregate 1). In Fig. 5, we can observe that this is realized with performances comparable to the ones obtained in Simulation 1. This is confirmed by the numerical indices in Table IV, where we observe that the MAPE is 0.54 % versus the 0.43 % obtained in Simulation 1, and the maximum APE reaches 5.12 %, not much higher than the 4.88 % achieved in Simulation 1. Simulation 4 aims at testing the capability of the MPC algorithm in adapting the control action to a different aggregate. Indeed, in this case, Aggregate 2 is simulated. Based also on the numerical indices in Table IV, we can state that the accuracy in tracking the reference signal is comparable to the ones obtained in Simulations 1 and 3.

## IV. CONCLUSIONS

In this paper, an adaptive MPC method has been proposed to modulate the power consumption of an aggregate of thermostatically controlled domestic EWHs, providing a DR service defined according to the rules of the Italian TSO. The approach is straightforward and only uses the measurements of the aggregated power consumption.

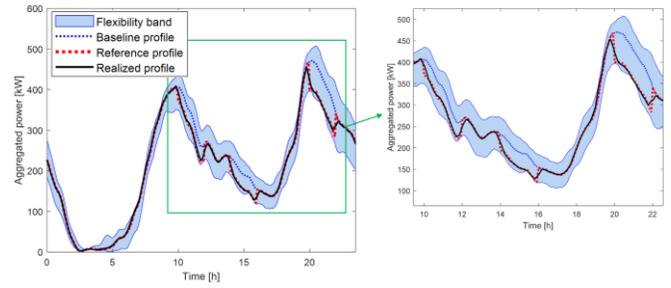

Fig. 5. Results of Simulation 3: power profiles.

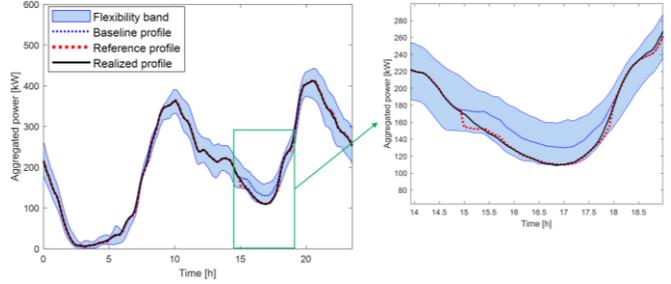

Fig. 6. Results of Simulation 4: power profiles.

The simulation analysis considers two aggregates composed of common commercial devices and proves that the DR service can be successfully provided by adopting the proposed method.